\documentclass[12pt]{article}
\usepackage{amssymb}
\usepackage{amsmath}

\def\cut{\hfil\break}
\def\beq{\begin{equation}}\def\eeq{\end{equation}}
\def\bea{\begin{eqnarray}}\def\eea{\end{eqnarray}}

\begin{document}
 
\title{Can gravity be added to Pilot Wave models?}
 
\author{Roman Sverdlov
\\Raman Research Institute,
\\C.V. Raman Avenue, Sadashivanagar, Bangalore -- 560080, India}
\date{September 30, 2010}
\maketitle
 
\begin{abstract}

\noindent The purpose of this paper is to come up with one of the many possible schemes of "adding" gravity to Pilot Wave models. 

\end{abstract}
 
\subsection*{1. Introduction}

One of the key problems related to quantum theory of gravity is well known. If gravitation is in fact a quantum field, then geometry, itself, undergoes quantum fluctuations. At the same time, in order to define quantum fluctuations of any field (including gravitational ones), we need to assume some kind of geometry of a system. Thus, we end up with a circular argument: we need to know geometry in order to define its quantum fluctuations. Beyond being circular, this argument is also contradictory: from path integral formalism we know that any quantity that fluctuates actually takes \emph{all possible} paths. Therefore, the existence of apriori geometry contradicts this assertion.

Similar questions arise in quantum mechanics in general, even if gravity did not exist. For example, in order for measuring apparatus to cause a "collapse" of the wave function, the former has to exist on the first place. So what caused a "collapse" of the measuring apparatus, if the latter have not yet came into existence? If there were other classical objects previously, what about the very first classical object? It is easy to see that this circle, in its logic, is similar to gravity. In one case we need classical object in order to produce classical object; in another case we need apriori geometry in order to produce the dynamics of geometry. Thus, the question of quantum theory of gravity is intrinsically linked to the one of the interpretation of quantum mechanics. 

There were some attempts in taking this approach, including spacetime quantum mechanics due to Hartle (\cite{Hartle}), generalization of Mensky's quantum corridors (Appendix of \cite{Thesis}),  quantum measure due to Sorkin (\cite{Sorkin}), my own version of preclusion principle (\cite{Preclusion}), etc. Most of these models assumed that gravitation is, in fact, one of the quantum fields and they attempted to find a way to cause it to "collapse" \emph{along with} other fields.  In this paper I will take a radically different approach. I will take a view, that was originally speculated by Dyson (see \cite{Dyson}), that gravity is striclty a classical field.  

As Dyson pointed out,  up to this day, there were no experiments that have conclusively proven the existence of gravitons or any other quantum behavior of gravity. Our only evidence of the latter comes from classical world. Thus, it is perfectly "legal" according to scientific method to assume that quantum gravity simply does not exist.  This, however, implies that the sources of the "classical" gravitational field have to be "classical" as well. This again brings us back to the question of "foundations of quantum mechanics": why are many-particle systems "classical", despite the fact that they consist of quantum mechanical particles? Thus, we have to \emph{first} answer this question and \emph{then} use the "classical" substances we produce as sources of gravitational field.

The exploration of flat-space Pilot Wave models is beyond the scope of this paper. Thus, we will simply take it for granted that we all agree on one of them (without specification which one). Furthermore, we will assume that the Pilot Wave model that we "agree upon" can be easilly generalized to the situation with \emph{classical} gravitational field. Again, we will assume that we already know how to do it, without explicitely writing how. Finally, we will claim that the "classical" objects that "move" according to one of the Pilot Wave models produce gravitational field. As we will soon see,  \emph{this part} is the most difficult one, despite the fact that it is "classical". 

In section 2 we will attempt to adress this issue within Everett's framework, and fail. We will then use this failure as a justification for introducing gravity into Bohmian framework in section 3. However, in Bohm's case, we will find that gravity will need to be modified in order to be produced by "non-conserved" energy momentum sources that Bohmian framework demands. Then, in section 4, we will propose one specific way of this modification. Finally, in section 5, we will have a qualitative discussion of the direction of future research that would possibly replace the proposal of section 4 with more natural-looking theories.

As a disclaimer, the specific modification to gravity that is proposed in this paper appeals to the concept of absolute time, which contradicts general relativistic covariance. In this sense, gravity is viewed as a classical field, separate from unalterable, apriori, geometry. This is justified by the fact that we are forced to break relativistic covariance regardless, since the sources of gravitational fields (which are "beables" of Pilot Wave models) are subject to superluminal signals. However, I still think that it would be better to couple non-relativistic Pilot Wave model to the version of gravity that completely respects relativity in its orthodoxy. That way, at least, we can claim that we have violated it to "the least possible extend". However, at this point I am not aware of ways of doing it. But this is definitely an important topic for the future research.

\subsection*{2. Everett + Gravity}

\subsection*{2.1 Everett's model in flat, non-relativistic, case: a review}

In order to understand Everett's model, it is important to remind ourselves that multiparticle system (which, in our case, is entire universe) can be viewed as a single point in multidimensional configuration space. The coordinates of that single point are the same as coordinates of all of the particles in the universe taken \emph{at the same time}. Thus, we have to introduce a "preferred" frame, and a "global" time $t$.  There is no "true" measuring aparatus; anything that we call by that name is really part of the system \footnote[1]{It is common among the scientists dealing with Everett's model to separate the "system" from the "environment". I do not agree with this view since this distinction is man-made and raises the same kinds of questions as the notion of "measuring apparatus" in Copenhagen approach. For this reason, as far as this paper is concerned, I will assume that \emph{everything} is part of the "system" and there is no such thing as the "environment"}. Thus, a single point in a configuration space will tell us the information about the entire universe, including an arrow in measuring aparatus pointing in some direction.

 Now, suppose the measurement is performed at some global time $t_1$ (by "measurement" we really mean the interaction of the particles of the apparatus with some other particles which, according to our view, stand on equal grounds). Suppose one of the outcomes of the measurement at $t_1$ is represented by a point $a$, and suppose another outcome is represented by a point $b$. Each $a$ and $b$ tells us about the coordinates of \emph{all} of the particles in the universe; this includes the particles that make up the arrow of measuring apparatus. Thus, we can assume that, among other things, point $a$ tells us that the arrow of measuring apparatus points in a direction $A$; on the other hand, point $b$ tells us that the arrow points in a direction $B$. 

Now, suppose there is another point in a configuration space, $c$. Furthermore, suppose that, within the time interval $t_1 < t < t_2$, the point $c$ is likely to be reached \emph{both} from $a$ \emph{and} from $b$. What this means is that, by "looking" at the configuration of particles represented by a point $c$, we can not deduce whether the arrow of measuring apparatus was pointing in a direction $A$ or a direction $B$. However, in light of the fact that the arrow is "big", it is likely to have interacted with a lot of other particles that bounce off of it. Thus, each of the direction it points into produces its \emph{own} set of irreversible changes; which does, in fact, enable us to deduce where the arrow was pointing at. Therefore, by contradiction, the above point $c$ is unlikely to exist. 

Furthermore, we can \emph{not} cheat and perform a \emph{new} experiment and hope that arrow would point in a direction $B$ the second time, if it pointed in the direction $A$ the first time. Even if it would, as a result of the fact that some time has passed, some irreversible changes have occured in the configuration of the rest of the particles simply due to thermal oscillations. Thus, the impact of the arrow on the \emph{new} configuration of particles can not possibly be the same as it would have been on the old one. The system is complex enough to "remember" not only \emph{where} the arrow pointed, but also \emph{when} it did. 

This implies that the wave function "splits into branches" that no longer overlap. This phenomenon is called \emph{decoherence}. This, of course, goes against our intuition in few-particle case. In these situations, the branches do, in fact, overlap once long enough time passes. This, however, is due to the fact that few-particle system is "small enough" so it only has impact on small number of particles, which, in fact, can be mimicked "by accident". The fact that branches no longer overlap is strictly a consequence of complexity of the system, which leads to a "classical" version of second law of thermodynamics. This is the ultimate reason why measuring devices are complex.

It is also important to note that, even in multiparticle state, each \emph{individual} particle can "overlap" with other particles if we think of it in terms of ordinary space. For example, while points $a$ and $b$ clearly disagree regarding the direction the arrow is pointing at, they might very well agree regarding the location of some of the other objects. The reason we say branches "don't overlap" is because, \emph{from the point of view of configuration space}, the coordinates of different particles are really coordinates of \emph{the same} point. Thus, if arrow points in different directions in points $a$ and $b$, this makes them, \emph{by definition}, separate, and anything else is irrelevent. 

Finally, it is important to mention that the above qualitative argument is not an addition to a theory. Instead, we are making a \emph{conjecture} that a wave function, which obeys Schrodinger's equation \emph{alone} undergoes decoherence. The verification of this conjecture is up to numeric simulations, which can be done without a single word of any qualitative argument. The only purpose of the above argument is to simply convince oneself that, contrary to one's intuition from few particle systems, the simulations will, in fact, predict the splitting of the branches. Likewise, that argument can be used to convince oneself that no mistake has been made in the simulations if they, in fact, produce the above result. 

Now, the key to collapse of wave function is this:  Since these branches no longer overlap, people living inside of each "branch" think that collapse of wave function have occured. In reality, however, that splitting into branches is simply a consequence of Schrodinger's equation (or some other evolution of states in case of quantum field theory). Thus, the latter dynamics has never been interrupted by "measurement" which means that it can, in fact, be viewed as classical. However, in order to "correctly" assess the dynamics of the wave function, we have to simulteneously look at \emph{all} of the branches and "add" them. Since we can not see other branches (due to lack of interaction with them) we have incomplete picture. Thus, the "incomplete" wave function that we see does, in fact, violate the evolution equation at a point of a "split". 

Again, we have not added anything "new" here. When we say we "don't see" something, what me mean is that the molecules of our eyes do not react to what happens in the other "universes". The fact that our eyes observe the collapse of wave function can also be predicted based on Schrodinger's equation, alone. Furthermore, computer simulations can also predict that the complexity of the system a particle interacts with is vital for collapse; in fact, if the computer is powerful enough we can do simulations to see just \emph{how} complex the system should be for collapse to occur. Furthermore, in light of the fact that we have appealed to irreversibility in our arguments, it might be guessed that temperature also contributes to the phenomenon. This, too, has been verified in the lab. 

\subsection*{2.2 Adding gravity to Everett's model: a failed attempt}

Now that we have reviewed the Everett's model it is time to do what we originally set out to do: introduce gravity into this model. Since the "complete" wave function obeys "classical" evolution equation that is never interrupted by measurement, we might as well view it as classical, which makes it fit to be a source of gravitational field. Of course, however, the wave function that we have is living in a configuration space (or, in case of quantum field theory, Fock space), while gravitational field is in ordinary space time. This means that we have to "project" the former into the latter. 

Let us now ask ourselves exactly \emph{what} do we want to project: should we project $\psi$, or $\vert \psi \vert^2$, or something else? To answer this question, we remind ourselves of our goal: we want to use whatever we project as a source of gravitational field. According to Bianchi identity, Einstein's equation can not be solved unless the energy-momentum tensor is conserved. We know that, according to Heizenberg's picture, operators obey "classical" equations. Thus, the fact that, classically, $\nabla^{\mu} T_{\mu \nu} = 0$ we know that the same identity holds if we replace $T_{\mu \nu}$ with an operator $\hat{T}_{\mu \nu}^{(H)}$, where "H" stands for "Heizenberg":
\beq \nabla^{\mu} \hat{T}_{\mu \nu}^{(H)} (t, \vec{x}) = 0 \eeq
where we have assumed that spacetime is curved, but we know the netric through some "classical" mechanism, which allowed us to use $\nabla^{\mu}$ as opposed to $\partial^{\mu}$.  Now, since in Heisenberg picture the states, themselves, do not evolve, the same conservation law also applies to $<\psi^{(H)} \vert \hat{T}_{\mu \nu}^{(H)} \vert \psi^{(H)}>$. Thus, we obtain
\beq \nabla^{\mu} < \psi^{(H)} \vert \hat{T}_{\mu \nu}^{(H)} (t, \vec{x}) \vert \psi^{(H)} > = 0 \eeq
Therefore, if we assume that the "infinities" (such as vacuum energy, renormalization issue, and so forth) have been taken care of, we can postulate a "classical" Einstein's equation to be 
\beq R_{\mu \nu} - \frac{1}{2} Rg_{\mu \nu} = < \psi^{(H)} \vert \hat{T}_{\mu \nu}^{(H)} (t, \vec{x}) \vert \psi^{(H)} > \eeq
which produces the desired "classical" dynamics for the gravitational field. For the Schrodinger's picture, we take $t$ dependence from $\hat{T}_{\mu \nu}$ and move it to $\vert \psi >$. At the same time, in light of the fact that time is absolute, $\vec{x}$ dependence has nothing to do with it. The latter simply refers to multiple degrees of freedom of our state \emph{at the same time}. Therefore, $\vec{x}$-dependence continues to sit inside $\hat{T}_{\mu \nu}$. Thus, the Schrodinger's version of gravity is 
\beq R_{\mu \nu} - \frac{1}{2} Rg_{\mu \nu} = < \psi^{(S)} (t) \vert \hat{T}_{\mu \nu}^{(S)} (\vec{x}) \vert \psi^{(S)} (t)> \eeq
Now let us see what happens in case of decoherence. We recall from the non-relativistic picture presented in previous chapter that in case of decoherence the wave function splits in non-overlapping branches. The generalization to our case is that, in \emph{Schrodinger's} picture, the state $\vert \psi^{(S)} (t)>$ splits as follows:
\beq \vert \psi^{(S)} (t) > = \sum_{k=1}^n \vert \psi_k^{(S)} (t) > \; ; \; t_1 < t < t_2 \eeq
where $\psi_k$ is \emph{part} of $\vert \psi >$ representing branch number $k$, and $t_1$ and $t_2$ are the timings of two subsequent measurements. Thus, the equation for gravity becomes
\beq R_{\mu \nu} - \frac{1}{2} Rg_{\mu \nu} \approx \sum_{i, j} < \psi_i ^{(S)} (t) \vert \hat{T}_{\mu \nu}^{(S)} (\vec{x}) \vert \psi_j^{(S)} (t)> \eeq
Based on decoherence, it is easy to see that $< \psi_i ^{(S)} (t) \vert \hat{T}_{\mu \nu}^{(S)} (\vec{x}) \vert \psi_j^{(S)} (t)> \; \approx 0$ whenever $i \neq j$. Thus, the above equation becomes
\beq R_{\mu \nu} - \frac{1}{2} Rg_{\mu \nu} \approx \sum_k < \psi_k ^{(S)} (t) \vert \hat{T}_{\mu \nu}^{(S)} (\vec{x}) \vert \psi_k^{(S)} (t)> \eeq
Now, physically speaking, $< \psi_k ^{(S)} (t) \vert \hat{T}_{\mu \nu}^{(S)} (\vec{x}) \vert \psi_k^{(S)} (t)>$ is the value of $T_{\mu \nu}$ measured inside of the universe number $k$. Thus, \emph{all} of the universes contribute to the metric! In other words, in our particular universe an arrow of a measuring apparatus is pointing in one, given, direction. But, according to our prediction, it should be attracted to all of the \emph{possible} arrows poining in all other \emph{possible} directions!  Common sense tells us that the amount of matter in our particular universe is much smaller than the amount of matter in all other possible universes. Thus, the gravitational field of the objects in our universe will be lost in much larger "white noise". 

One might attempt to solve this problem by saying that there are several different gravitational fields for each universe, and the universe number $k$ posesses a gravitational field $g^k_{\mu \nu}$, which obeys an equation 
\beq R^k_{\mu \nu} - \frac{1}{2} R^k g^k_{\mu \nu} \approx \; < \psi_k ^{(S)} (t) \vert \hat{T}_{\mu \nu}^{(S)} (\vec{x}) \vert \psi_k^{(S)} (t)> \eeq
In the above expression we have left the approximation sign for a reason. Any attempt to turn it into exact equality inevitably leads to the following problems:

a) We can not rigurously define a boundary between different universes; thus, we can not come up with an \emph{exact} definition of $\vert \psi_k >$

b) Every time a "split" occurs, we introduce more and more versions of $g_{\mu \nu}$. This "jump" on the number of variables does not have a well defined location in time. 

c) During the "split", the \emph{total} energy-momentum is concerved. The latter includes the sum of energy momentum over \emph{all} of the branches after the "split". Thus, if we are to look at only \emph{one} of the branch, we might find that the energy momentum tensor will be no longer conserved. If that is the case, it can be easilly shown, by using Bianchi Identities, that Einstein's equation will no longer have solution.

Issues "a" and "b" can be adressed if we replace come up with more "precise" definition of a universe. In particular, instead of viewing universe as a "branch" of a wave function, we can view it as one single state (or, equivalently, one single point in a Fock space). That state evolves under the "guidence" of $ \vert \psi >$ and then, when $\vert \psi >$ splits, it is forced to "select" the branch it will go into. The source of gravitational field will be that single state, as opposed to the whole branch, thus the issue "a" is adressed. Furthermore, since that state is the only source of gravitational field, and it did not move into "other branches", the latter does not gravitate. This means that there is only \emph{one} version of $g_{\mu \nu}$, which adresses the issue "b".

However, the issue "c" in this proposal is \emph{not} adressed; in fact, it has possibly been made worse. If we are dealing with a "large" branch, then there is at least \emph{some} hope that, at large enough scale, different violations of energy-momentum tensor could have been averaged out to approximately $0$. When we introduce one single state, that "averaging out" is less likely. While it would still happen, it would require even larger scale. 

Nevertheless, since we are dealing with an exact science, we don't care as much whether or not the issue "c" was "made worse" or not. The fact is that issue "c" has been there both in case of a "branch" as well as in case of a "single state". Issues "a" and "b", on the other hand, went away for the case of single state. This clearly means that the picture of single, evolving, state is favorable. In order to adress issue "c", we will have to come up with a modification for classical gravity that will allow the "modified" Einstein's equation to have a solution for non-conserved case. This is, in fact, the main point of section 4.

It is important to point out that the existance of that single, evolving, state is the main feature that distinguishes Bohm's view from Everett's. Thus, we have basically shown that Everett's model has miserably failed, while Bohm's model, while still not perfect, at least makes things somewhat better. This should not be viewed as the ultimate rebuttal of Everett. In fact, other models of quantum gravity, including \cite{Hartle}, \cite{Sorkin}, and Appendix A of \cite{Thesis}, are more compatible with Everett's view. Rather, what I have shown is that the \emph{specific} way of introducing gravity that is being proposed in this paper. The discussion of other models, however, is beyond our scope. 

\subsection*{3.  Bohm's model}

\subsection*{3.1  Overview of flat cases}

Consider, first, non-relativistic quantum mechanics with one single particle. According to Bohm, we have two \emph{separate} substances, a particle and a wave, living in a multidimensional configuration space. The wave evolves according to Schrodinger's equation, 
\beq - \frac{1}{2m} \frac{\partial^2 \psi}{\partial t^2} + V \psi = E \psi ,\eeq
while a particle moves according to \emph{guidence equation},
\beq \frac{d \vec{x} (t)}{dt} = \frac{1}{m} \; Im \; \vec{\nabla} \; ln \; \psi \eeq
 The quidence equation postulates one-way influence of the wave on the particle and implies that the behavior of the wave completely determines the behavior of a particle, if we know its initial location. At the same time, the particle has no influence on the wave: the behavior of the wave is completely determined by Schrodinger's equation (or some quantum field theory version of it), and does not take into account the behavior of a particle.

The above equation is deterministic. Thus, if we know the wave function, as well as the position of a particle at a given point in time, we will know its exact location at any future point in time. This, however, leaves a room for \emph{classical} version of probabilities. If we don't know the exact initial conditions then we would be able to quantify the degree of our "human ignorance" by "probabilities". Thus, strictly speaking, a "probability" will be different for each person, depending on their level of "ignorance". In particular, if we define a "velocity field" $\vec{v}$ by
\beq \vec{v} (\vec{x}) = \frac{1}{m} \; Im \; \vec{\nabla} \; ln \; \psi \eeq
then the guidence equation can be rewritten as $d \vec{x} / dt = \vec{v} (\vec{x})$. This will imply that, by continuity equation, the probability will evolve as 
\beq \frac{\partial \rho}{\partial t} = \vec{\nabla} \cdot (\rho \vec{v}) \eeq
At the same time, it is easy to show from Schrodinger's equation that 
\beq \frac{\partial \vert \psi \vert^2}{\partial t} = \vec{\nabla}\cdot  (\vert \psi \vert^2 \vec{v}) \eeq
Thus, if $\rho = \vert \psi \vert^2$ at \emph{some} point in time, this equality will continue to hold later on; this means that $\rho = \vert \psi \vert^2$ is an \emph{equilibrium point} of the theory. However, the inspection of above equations shows that they satisfy time reversal symmetry. Therefore, just like the system can not \emph{leave} the equilitbrium, it can not \emph{enter} the equilibrium either. The philosophy of determinism tells us that initial conditions can be anything we like, which means that, with absolute certainty, we were \emph{not} at equilibrium, initially.  Therefore, we never will be. 

Nevertheless, while $\rho = \vert \psi \vert^2$ can not be obtained exactly, it can approximately hold on a \emph{coarse grained} level. That is, if $N_{\delta} (\vec{x})$ is a $\delta$-neighborhood of a point $x$, then, up to some approximation, 
\beq \int_{N_{\delta} (\vec{x})} d^3 x' \; \rho (\vec{x}') \approx \int_{N_{\delta} (\vec{x})} d^3 x' \; \vert \psi (\vec{x}') \vert^2 \eeq
In this case it is, in fact, much easier to "enter" the above situation than to "leave" it, which almost guarantees that the above situation will occur once enough time has passed. This is a phenomenon of a \emph{classical} version of the increase of entropy. If we take a blue sand and a red one, then, on a level of \emph{points}, the red pieces of sand will remain red and the blue ones will remain blue, no matter how much we mix them. But, on a level of "neighborhoods", the colors do in fact "mix": at first most of the neighborhoods contained only one color, but later on they contain both colors, and this process is very unlikely to be reversed. 

In light of the fact that our senses are imprecise, whenever we ask ourselves whether or not a particle can be found at a point we are, instead, asking about a neighborhood of a point. That is why, at equilibrium, we do get an answer $\rho = \vert \psi \vert^2$. As stated previously, $\rho$ is a probability in a \emph{classical} sense, which means that it is only related to the ignorance of the observer. Thus, from the point of view of different observers, the time it takes to reach the equilibrium is different. Yet, they all agree that the "equilibrium" will be reached, since they are all "ignorant" to some extend. 

Now, suppose we replace a single particle with a configuration of many of them. In this case, according to the previous chapter, the wave function "splits" into branches at some point. Bohm agrees with Everett in this respect. As a result of this splitting, the point particle is forced to go into \emph{one} of these branches. Due to the fact that the value of $\vert \psi \vert^2$ between the branches is nearly $0$, and the trajectory of a particle is continuous, the particle is forced to stay inside the branch it went into. Thus, for example, if it went into the branch $k$ (which we will call $B_k$), then the probability of finding the particle within that branch is 
\beq \rho (\vec{x}) \approx \frac{ \vert \psi (\vec{x}) \vert^2}{\int d^3 x' \; \vert \psi (\vec{x}') \vert^2} \; , \; x \in B_k \eeq
while the probability of finding it within any other branch is 
\beq \rho (\vec{x}) \approx 0 \; , \; x \in B_l \; , \; l \neq k \eeq
Since the above picture is in a configuration space, a single point represents the location of \emph{all} particles. Just like in Everett's case, each of the "branches" represents the arrow of measuring apparatus pointing in different direction. However, this time, we identify our "universe" with a particle rather than a branch. Thus, there is only \emph{one} universe, and the arrow points in only \emph{one} direction.  In the context of Bohm's framework, this phenomenon is called "effective collapse". We have an appearance of collapse, while in reality there was none: both Schrodinger's equation as well as quidence equation, continued to hold at all times.

One can adapt a "hybrid" point of view and claim that the universe is a branch rather than a particle but, at the same time, a particle picks out a "preferred" branch. In other words, we have one single universe, which is identified with a branch that was "selected" by the particle. This view is probably the reason why sometimes that particle is referred to as "hidden variable". This point of view can be justified by the fact that there is no "splitting" between the vicinity of the particle and the rest of the branch, which means that the specific location of the particle will not be "recorded". Of course, in light of determinism of a model, we would be able to deduce what it used to be \emph{if} we know its present location. But, from what we said before, we only know $N_{\delta} (\vec{x})$ as opposed to $\vec{x}$, and the latter is not sufficient to decude the past of the particle. Thus, if we stick to the point of view that the "reality" is something that we can "remember", then the reality will be identified with a branch, and the particle will be identified with a "hidden variable". 

However, this point of view is unacceptable if we intend to introduce gravity. As mentioned at the end of the last section, the definition of branch is too imprecise for it to be viewed as a source of gravitational field. Therefore, if we are to have a precise theory, we have a choice between claiminig that \emph{all the branches} gravitate, or claiming that none of them do. The former claim got us into trouble in the last section. If we will instead stick to the latter claim, then we are forced to identify the particle, itself, as a sole source of gravitational field. This means that the particle should, in fact, be viewed as reality. In fact, the gravitational field might, in fact, allow us to "record" where the particle has been located inside the branch, thus countering the point made in the previous paragraph. 

If we generalize Bohm's Pilot Wave model to quantum field theory, then the "wave" in a Fock space is simply a set of probability amplitudes of different quantum states. A "particle" in a Fock space is one specific quantum state (which, from the perspective of \emph{ordinary} space consists of many particles) that is being realized (in the context of position beables this state can be written in terms of position of particles which, in "flat" case, can be easilly obtained as a Fourier transform of states in momentum space). Thus, since a "particle" only exists at \emph{one single} point, there is only \emph{one} configuration of particles that takes place at any given point in time. Despite this fact, we still have a set of \emph{non-zero} probability amplitudes of all other "possible" configurations. The latter are not true "probabilities" since the dynamics of the "actual" configuration is deterministic, thanks to Pilot Wave model. The notion of "probability amplitude" is simply a generalizatin of a notion of "wave" onto configuration space. The "actual" configuration evolves based on the equation that takes into account all of the "possible" ones and their amplitudes.

There is one major problem, however, when it comes to generalizing from configuration space to Fock space. As we have seen earlier, the key to determinism is the fact that the guidence equation is differentiable. But, due to the fact that creation and annihilation of particles is not a differentiable process, the latter can not be modeled in a deterministic way unless we do some manipulations. There were several proposals as to how to go around this issue. Since the purpose of this paper is to introduce gravity into \emph{generic} Pilot Wave model, as opposed to sticking with each particular one, we will simply list a sample of the approaches without choosing any of them over others:

a) Field beables (\cite{minimalist}). According to this model, the fields replace particles as beables. For example, if we have a scalar field, then we can view its \emph{second} quantization in terms of a \emph{first} quantization of a lattice which lives in 4+1 dimensional spacetime $(t; \phi, x, y, z)$ and oscillates in $\phi$ direction, while the $x$, $y$, and $z$ coordinates of its points are fixed. Thus, the dimensionality of a configuration space is fixed, since the lattice points can neither be created nor destroyed. This allows us to apply Bohm's model to this system in deterministic way. The "point" in a configuration space is a "shape" of a lattice; or, in other words,  classical spin 0 field. Hence, the notion of "field beable". This is easilly generalizeable from spin 0 to spin 1 case. However, this can not be generalized to fermions, due to the Grassmannian nature of these fields. Nevertheless, one can claim that we have never "seen" fermions anyway: we only "see" the photons that come out of electron and this makes us conclude that there is, in fact, an electron. From this point of view (which Struve and Westman referred to as \emph{minimalist} one), the only thing we should be able to predict is the behavior of bosons, which this model allows us to do.

b) Dirac see (\cite{collin}). This model appeals to the original concept proposed by Dirac: fermions fill all of the negative energy levels of "dirac see" and when they absorb a photon they "jump" to one of the positive energy states. This creates a "hole" in the Dirac see, and this "hole" acts like a positron. Thus, no pair creation has occured: we continue to have the same, fixed, number of electrons, without any positrons. True: there was a discrete jump in energy level. But this does not cause us too much trouble since we do not view energy as a beable; we are using position as such. Thus, as long as the position of the electrons changes in a continuous fashion, we do have a room for a deterministic model, despite the "jumps" in energy level.

c) Discrete jumps (\cite{jumps}). According to this model, there are finite time intervals, interrupted by random "jumps". During each of the intervals, the number of particles is fixed, and they evolve according to some deterministic equation. The timing of a jump, however, is random. If, at a time $t$, the system is at a state $\vert e >$, then the probability that, at a time $t+dt$, the system will be found at a state $\vert e'>$ is given by $\sigma (\vert e>, \vert e' >) dt$, where
\beq \sigma (\vert e >, \vert e'>) = \frac{(<\psi \vert e> <e \vert H \vert e'> <e' \vert \psi>)^{\dagger}}{\vert <\psi \vert e > \vert^2} \eeq
provided that one can not move from $\vert e>$ to $\vert e'>$ without making a discrete jump. In the above expression, $x^{\dagger}$ is $x$ for $x>0$ and $0$ for $x<0$; thus, the expression is positive. It can be shown that this kind of stochastic process does, in fact, produce the desired probability density $\vert < \psi \vert e > \vert^2$. 

d) Effectivity/Visibility of particles (\cite{nikolicb}, \cite{nikolicf} and \cite{Italy}). According to these models, particles do not trully get created or destroyed. Instead, they "hide" by having their effectivity/visibility approach $0$. On the other hand, they un-hide if their effectivity/visibilty is close to $1$. Since effectivity/visibility varies in continuous fashion, this allows the process of their creation/annihilation to be continuous as well. According to the "visibility" version of the argument (\cite{Italy}) most of the time the visibility is close to $0$ or $1$, while transition periods are still finite, but very small. According to the "effectivity" version of the argument (\cite{nikolicb} and \cite{nikolicf}), this does not have to be the case. Both models, however, imply differentiable and, therefore, deterministic, behavior. 

e) The "big computer" (\cite{Czech}). In this model we imagine a "classical" universe in which a "computer" has been built that was programmed to "simulate" Fock space on its screen. Since any kind of "program" computer might have has to do with the interaction of its particles, what this means is that, due to clever enough setup, the latter produced the \emph{appearance} of quantum phenomena. At the same time, since the computer was built in "classical" world, everything is "classical" by default. We can use this scheme and refer to anything we "don't like" as "quantum mechanical". In particular, we "don't like" configuration space, we "don't like" lack of determinism, and we "don't like" discrete jumps. We thus come up with a model of continuous and deterministic processes in ordinary space that involve fixed number of particles which, due to the clever enough setup, lead us to believe in configuration space, particle creation/annihilation, and so forth. 

\subsection*{3.2  Adding gravity to Pilot Wave model: qualitative discussion}

We have established that introducing gravity into the Everett's framework might create a problem due to the fact that different "universes" interact gravitationally. In Bohm's case, this problem is avoided since we only have one single universe. However, we now have a different problem. In light of the fact that we view gravitational field as "classical", it has to be an exact solution to Einstein's equation. From Bianchi identities it can be easilly shown that Einstein's equation only has a solution if energy-momentum tensor is conserved. If, however, we associate the latter with the motion of particles, then it is not conserved due to their acceleration. 

Thus, if we insist on leaving Einstein's equation in its original form, we have insist that wave \emph{does} have energy-momentum and then come up with a dynamics in which a particle "steals" energy-momentum from a wave and "concentrates" it around itself. Since the energy momentum of a wave is defined in terms of its dynamics, a particle has to have an influence on the dynamics of a wave in order for the "stealing" not to violate conservation laws. Thus, none of the existing Pilot Wave models are fit, since in all of them only wave has an influence on a particle, and not the other way around. 

Another way to see this is to remind ourselves that, based on the proof of Noether's theorem, we know that the conservation of energy-momentum is a consequence of the existence of the Lagrangian. Now,  in order for a Lagrangian to imply the influence of a wave on a particle, it would have to couple the two. The coupling, however, will \emph{also} imply the influence of a particle on a wave, which is absent from Pilot Wave models. Thus, while it might seem that a "mechanical" view of Pilot Wave model proposed in \cite{Czech} might be a step in the desired direction, we know that this is doomed from the start simply because some of the "mechanisms" proposed in that paper involve one-way interactions. 

All of the one-way interactions have to be replaced with two-way ones if we are to have any hope of restoring energy-momentum conservation. At the same time, if we do the latter without any kind of "cheating", we can not guarantee that a particle will, in fact, "steal" most of the energy from the wave.  In fact, since \emph{up to some approximation}, we expect Bohm's view to hold, we will probably see that a particle steals only \emph{a little bit} of energy, which is just enough for the particle to "move" under wave's "guidence"; at the same time, the wave continues to posess most of the energy. But, in this case, we are back to the problem we encountered with Everett's view: most of the energy-momentum will be dominated by "parallel universes" that have not been physically realized, which means that gravitational "white noise" will far outweigh any of the gravity of the "observed" objects. 

In light of these issues, we propose a different route. We will attempt to \emph{modify} the "classical" theory of gravity in such a way that it has solutions for non-conserved energy momentum. If we succeed, then we will claim, as proposed earlier, than the energy of the wave is exactly $0$, and the particle obtains energy-momentum "from nowhere" when it tries to move under the wave's guidence. We will then use that non-conserved energy momentum as a source of \emph{modified} classical gravitational field. The "modification" has to be done in such a way that the solution always exists. It is important to stress that this problem can be stated in purely "classical" terms: we want to come up with Einstein's equation that has a solution for non-conserved energy-momentum \emph{regardless} of the source of non-conservation. Its solution, however, will ultimately enable us to solve "quantum mechanical" problem of introducing gravity into the quantum world.

Interestingly enough, this question has "classical" interest, independently of quantum mechanics. Our intuition tells us that \emph{first} some fields (that include both gravity \emph{and} other things, such as electrodynamics) cause particle to move, and \emph{then} its motion has an effect on gravity. So \emph{what if} we were to change some non-gravitational fields (such as electromagnetic one) in such a way that it would no longer conserve energy momentum? Or \emph{what if} someone "outside the universe" were to move particle "by hand"? What would happen to gravity in these cases? Our cause and effect intuition tells us that the behavior of gravity has to continue to be well defined \emph{even then}, and yet the "classical" physics tells us otherwise!

 Of course, the violation of energy-momentum conservation contradicts our intuition, too. But this issue can be easilly adressed by slightly changing the semantics. What we were referring to as energy-momentum up till now was a tensor $T_{\mu \nu}$. We can now claim that the latter is \emph{not} a "true" energy momentum; we only used these words for our convenience. The "true" energy-momentum is, instead, $R_{\mu \nu} - \frac{1}{2} Rg_{\mu \nu}$, which continues to be conserved by Bianchi identity. At the same time, the latter is no longer a part of Einstein's equation. Thus, our modification of gravity implies that the energy-momentum (which is still conserved) is no longer a source of gravity but rather a mathematical construct that plays no role.  On the other hand, the source of gravity is \emph{something else} (which we simply "happened" to call $T_{\mu \nu}$), which is not conserved. On a larger scale, the two happen to average out to approximately the same thing which is what lead us to \emph{wrongly} confuse them. 

\subsection*{4. Modification of gravity: one explicit example}

We will now propose one specific way of modifying gravity in order to encorporate Pilot Wave model. Our goal is to adjust to \emph{generic} Pilot Wave model, so we will not specify which one we are dealing with. Instead, we will look at the commonalities of the Pilot Wave models we have listed and write the "general" one. If our beables are particles, we would like to make sure they don't create black holes. Thus, we should postulate a "cloud" of $T_{\mu \nu}$ spread "around", instead of it being concentrated at a point. The specific way of doing it is up to the model. The acceleration of the particles implies that $T_{\mu \nu}$ is not conserved. 

On the other hand, if we have a field beables then, again, we can use common prescription to define $T_{\mu \nu}$. However, despite the fact that we use the same equation as we do in a "trully" classical case, $T_{\mu \nu}$ will not be conserved. The reason for this is that its conservation is a consequence of wave equation. But, as we mentioned earlier, second quantization of a scalar field can be modelled as a first quantization of the oscillating lattice. In the configuration case, the latter is a particle, not a wave. Thus, the "field beable" is also a "particle". This means it does not obey wave equation; instead, it obeys \emph{guidence} equation that is not related to the latter. Since the conservation is a consequence of \emph{wave} equation, we can no longer assume that $T_{\mu \nu}$ is conserved. 

Thus, the common feature of both models is a presence of some non-conserved $T_{\mu \nu}$, which we treat as a beable. Thus, for any specific Pilot Wave model, we are to "convert" it into the language of dynamics of $T_{\mu \nu}$; the "conversion" is specific to the model at hand (for example, in case of point particles, it might involve giving them a size). Furthermore, each of the specific Pilot Wave models is to be adopted to the case of a (classically) varying metric $g_{\mu \nu}$. In light of non-local nature of Pilot Wave models, the dependence might be on \emph{global} behavior of $g_{\mu \nu}$, which, on a manifold $\cal M$, will be denoted as $g_{\mu \nu} ({\cal M})$ (to distinguish it from the local one which we will call either $g_{\mu \nu}$ or $g_{\mu \nu} (x^{\rho})$). Thus, the generic Pilot Wave model is 
\beq \frac{\partial \vert \psi > }{\partial t} = H(\vert \psi >, g_{\rho \sigma} ({\cal M})) \; ; \; \frac{\partial T_{\mu \nu}}{\partial t} = V_{\mu \nu} ( \vert \psi > , g_{\rho \sigma} ({\cal M}), x^{\eta}) \eeq
where the use of $g_{\mu \nu} ({\cal M})$ is to remind us that the evolution depends on a \emph{global} behavior of a metric $g_{\mu \nu}$ throughout the entire manifold $\cal M$, \emph{as opposed to} at a local neighborhood of a point. 

Our remaining, and main, challenge is to find dynamics for $g_{\mu \nu}$ based on $T_{\mu \nu}$. As we have said previously, in light of the fact that $T_{\mu \nu}$ is not conserved, we can not expect Einstein's equation to be exactly satisfied. What we \emph{can} do, however, is to do "trials and errors" in which we vary $g_{\mu \nu}$ in "absolute" time $t$ and asking ourselves whether or not we are getting any closer to the desired approximate result. If we do, then we continue to vary $g_{\mu \nu}$ in the same way we used to. Otherwise, we vary it in the opposite direction. The purpose of the rest of this chapter is to come up with "deterministic" mechanism of performing seemingly-random trials and errors.

Qualitatively speaking, the "scheme" is the following. There are some "short" time intervals, separated by the long ones. The "short" time intervals are local. That is, if we use some global coordinates to define a point in "space" (as opposed to spacetime), then the timing of the "short" intervals are unique to every local neighborhood in space, and their timing does not match for any two different neighborhoods. Now, whenever a point $x^{\mu}$ is subjected to the "short" time interval, the random variation of $g_{\mu \nu}$ is produced, which might \emph{either} bring $R_{\mu \nu} - \frac{1}{2} Rg_{\mu \nu}$ closer to the desired value, \emph{or} further apart. Then, during the subsequent "long" time interval, the variation of $g_{\mu \nu}$ continues in the same direction if the "short" time interval produced desired result, or it goes in the opposite direction otherwise. 

Therefore, during the "long" time interval we are biased in favor of the "right" direction; during the "short" time interval we are not. This means that as long as the "long" time interval is significantly longer than the "short" one, the "right" direction of change of $g_{\mu \nu}$ will dominate. At the same time, however, the "long" time interval should not be \emph{too} long, either. After all, if we go "too far" in the "right" direction, we might pass the extremum point and then the right direction will be a "wrong" one. Thus, we need some "short" intervals along the way to correct us. Adjusting the "correct" length of the "long" and "short" intervals is ultimately up to the correct choice of constants in the equations that follow. This possibly involves numerical analysis that is beyond the scope of this paper. For the sake of this preliminary work, we will simply assume that the correct choice of constants can be made. 

Now, in order to formally separate "short" and "long" intervals, we will introduce a "random generator" $\chi$ which evolves according to deterministic equation, 
\beq \nabla^{\mu} \nabla_{\mu} \chi + k \chi = 0 \eeq
We assume that our universe has compact topology, which is compatible with flat metric. For example, it might be a 3-torus in space, that stays unchanged with time. Of course, as a result of gravity the curvature will be "added" to it; but the topology itself does not force it to be there. Now, as a result of compact topology the $\chi$ field "bounces" back and forth and sometimes produces resonances. I propose that locations and durations of these "resonances" define the "short" time intervals. 

In order to clearly separate the "short" time intervals from the "long" ones, we would like to use $f(\chi)$ as opposed to $\chi$ itself where $f$ is a real valued function on a real line. The specific definition of $f$ is not important, as long as it is some form of differentiable approximation to step function, which would separate the values of $\chi$ "below" the "step" from the values of $\chi$ "above" one. For definiteness, we will define $f$ as
\beq f(\chi) = \frac{1}{2} + \frac{2}{\pi} tan^{-1} (n(\chi -\chi_0)) \eeq
where $n$ and $\chi_0$ are constants. It is assumed that $n$ is very large, but finite. Thus, we assume that the "effective" theory is a reality: $f$ is not a step function, but a very good approximation to one. It is easy to see that the larger $\chi_0$ is, the shorter is the "short" interval, and the longer is the "long" one. Thus, $\chi_0$ should be chosen in such a way that the intervals have the desired length. The selection of $\chi_0$ should also take into account the global topology of the universe. 

Now, as we have mentioned earlier, during each of the "short" intervals, $g_{\mu \nu}$ should undergo "random" change. Since we intend our theory to be deterministic, we would like to introduce a tensor-valued "random generator" $A_{\mu \nu}$ that evolves according to the equation
\beq \nabla^{\rho} A_{\mu \nu} \nabla_{\rho} A_{\mu \nu} + k A_{\mu \nu} = 0 \eeq
 and try to come up with a dynamics of $g_{\mu \nu}$ that would, during the "short" intervals, be approximated as 
\beq \frac{\partial^3 g_{\mu \nu}}{\partial t^3} \approx aA_{\mu \nu} f(\lambda) \eeq
where $a$ is some constant. The approximation sign has been used because we would like to have the same dynamics both for the short \emph{and} long intervals. Thus, the fact that the above dynamics holds true only for the short ones means that it is not complete; the complete equation will be introduced shortly. In the above expression, we have used $\partial_0$ instead of $\nabla_0$ in order to avoid the conflict with well known identity
\beq \nabla_0 g_{\mu \nu} =0 \eeq
The violation of general relativistic covariance is justified based on the fact that the covariance has \emph{already} been violated when Pilot Wave model has been introduced. Finally, we have used \emph{third} time derivative for $g_{\mu \nu}$ because the \emph{second} one contributes to curvature, which is the ultimate goal of the theory. Thus, the "derivative of a curvature" (or, in other words, a third derivative of a metric) has to be used for a dynamics. It is easy to see that the space derivatives will vary as a consequence of the time ones. Thus, we take into account both time \emph{and} space derivatives when we determine whether or not the "random" change during the "short" interval is "benefitial". 

Now, we need a mechanism of "remembering" the nature of random change as well as whether or not it was "benefitial". Then, during the "long" interval we will use that "memory" in order to determine the subsequent evolution of metric. Since the evolution of $A_{\mu \nu}$ has already been defined, we need a separate field, $B_{\mu \nu}$, in order to store that "memory". In other words, the complete evolution equation for $g_{\mu \nu}$ is
\beq \frac{\partial^3 g_{\mu \nu}}{\partial t^3} = aA_{\mu \nu} f(\chi) + bB_{\mu \nu} \eeq
During the "short" time interval, $f(\chi) \approx 1$ and, therefore, both $A_{\mu \nu}$ and $B_{\mu \nu}$ contribute. It is assumed, however, that $a >>b$, which is why $A_{\mu \nu}$ dominates. However, during the "long" time intervals, $f(\chi)$ is so close to $0$ that $a f(\chi) << b$, despite the fact that $a >> b$. That is why during the "long" intervals $B_{\mu \nu}$ dominates. 

Now, in order for $B_{\mu \nu}$ to serve a purpose of "memory", it has to evolve \emph{during} the "short" interval, and then retain the state it has evolved into during the "long" one. In other words, during the "short" interval, the role of $B_{\mu \nu}$ on a dynamics is insignificant \emph{but} its internal evolution is \emph{crucial}. Now, if $B_{\mu \nu}$ "sees" that $g_{\mu \nu}$ evolves in "unfavorable" direction during the "short" interval, it should try to attain the value of $- \lambda A_{\mu \nu}$; or, if it sees that $g_{\mu \nu}$ evolves "favorably", it should attain the value of $+ \lambda A_{\mu \nu}$.

Let us now try to define what we mean by "favorable" and "unfavorable" directions. We will write the "desired" results by using the word "want" above the equal signs or inequalities (for example, $=^{\rm want}$, $\approx^{\rm want}$, $<^{\rm want}$, $>^{\rm want}$, etc). Thus, our goal is for the field $g_{\mu \nu}$ to approximately obey Einstein's equation, 
\beq R_{\mu \nu} - \frac{1}{2} R g_{\mu \nu} - T_{\mu \nu} \approx^{\rm want} 0 \eeq
We will rewrite it in a differential form,
\beq \frac{\delta (\sqrt{-g} g^{\mu \nu}R_{\mu \nu} - g^{\mu \nu}T_{\mu \nu})}{\delta g^{\mu \nu}} \approx 0 \eeq
Now, the way we verify this condition is by introducing another tensor field, $h_{\mu \nu}$ that undergoes small oscillations, 
\beq \nabla^{\rho} \nabla_{\rho} h_{\mu \nu} = 0 \eeq
 and we will define $R^{g+h}_{\mu \nu}$ to be Riemann curvature tensor based on the metric $g_{\mu \nu} + h_{\mu \nu}$, while $R^g_{\mu \nu}$ is identical to $R_{\mu \nu}$. Now we introduce the function $\eta$ that evolves according to the equation
\beq \frac{\partial \eta}{\partial t} = - \alpha \eta + exp \Big( i \beta \Big(\sqrt{- det (g_{\mu \nu} + h_{\mu \nu})} \Big( (g^{\mu \nu} - h^{\mu \nu}) R^{g+h}_{\mu \nu} - T_{\mu \nu} \Big) \Big) \Big) \eeq
Due to the $- \alpha \eta$ term, the contributions from distant enough past to $\eta$ died off. Thus, if $\alpha$ is \emph{very} large, then the value of $\eta$ will record the interference caused by the "oscillations" of $h$ within a near vicinity in time. Now, it is easy to see that if we are closer to the "desired" configuration of the metric, it means that we are closer to equilibrium. Thus, the above interference will be "constructive". The further away we are from the equilibrium, the bigger the "disractive" interference we would get. 

Now, we would like to try to assess whether we are moving "towards" equilibrium or away from it. In order to do that, we will introduce another parameter, $\eta'$, that operates on slightly larger time scale than $\eta$. This can be accomplished by replacing $\alpha$ with $\alpha'< \alpha$, which still a very large number: $\alpha > \alpha' >> 1$: 
\beq \frac{\partial \eta'}{\partial t} = - \alpha' \eta' + exp \Big( i \beta \Big(\sqrt{- det (g_{\mu \nu} + h_{\mu \nu})} \Big( (g^{\mu \nu} - h^{\mu \nu}) R^{g+h}_{\mu \nu} - T_{\mu \nu} \Big) \Big) \Big) \eeq
Thus, $\eta'$ goes slightly "further back" into the past than $\eta$ does. If we are going towards equilibrium, then the level of interference decreases, which means that $\eta$ will have less interference than $\eta'$. On the other hand, if we are moving away from equilibrium, then the situation is the opposite. Now, suppose that we neither go towards the equilibrium nor away from it. In this case, based on the dimensional analysis, $\alpha \eta = \alpha' \eta'$. Intuitively, $\eta'$ had "more time" to build itself from the ingredients with the same degree of interference. Therefore, the ultimate criteria for whether or not we are moving in a favorable direction is the value of $\alpha \eta - \alpha' \eta'$: it is positive if we are moving in a favorable direction, and negative otherwise. 

Now, according to our "program", we are evaluating the desirability/undesirability of a change only during a "short" time intervals. We record the result in $B$ (which plays no active role during the "short" interval) and then the value of the latter will be crucial for the dynamics during the subsequent "long" interval. During the latter period of time we no longer need to look at the "desirabliity" of a change: until the next "short interval" takes place, we trust the information that we have obtained from the previous one.  Therefore, the above "diserability" criteria should be used \emph{only} during the "short" intervals; or, in other words, $\chi > \chi_0$. 

As we said before, we have replaced the latter inequality with a condition involving differentiable approximation to step function, $f (\chi)$. Thus, the "short" intervals are defined by $f (\chi - \chi_0) \approx 1$. Therefore, in order for $B_{\mu \nu}$ to become $+ \lambda A_{\mu \nu}$ two conditions have to be met: one is that $f(\chi - \chi_0) \approx 1$, and the other one is the above inequality. In other words, we can sum it up as
 \beq  \alpha \eta - \alpha' \eta' >0 \wedge f(\chi - \chi_0) \approx 1 \Rightarrow^{\rm want} [B_{\mu \nu} \rightarrow + \lambda A_{\mu \nu}] \eeq
Similarly, in order for $B_{\mu \nu}$ to become $- \lambda A_{\mu \nu}$ also two conditions have to be met. One of the two is the same as in previous case, $f (\chi - \chi_0) \approx 1$. The other one, on the other hand, is an inequality with flipped sign. Thus, 
 \beq \alpha \eta - \alpha' \eta' <0 \wedge f(\chi - \chi_0) \approx 1 \Rightarrow^{\rm want} [B_{\mu \nu} \rightarrow - \lambda A_{\mu \nu} ] \eeq
We would now like to summarize the above two equations in a single one. We have already introduced a differentiable approximation to step function, $f(x)$, that is approximately equal to $0$ for $x<0$ and $1$ for $x>0$. Thus, if we take $1 - 2 f(x)$, we will, in fact, get $+1$ for $x<0$ and $-1$ for $x > 0$, as desired. Thus, we substitute the above two expressions with 
\beq  f(\chi - \chi_0) \approx 1 \Rightarrow^{\rm want} \Big[ B_{\mu \nu} \rightarrow  \lambda A_{\mu \nu} (1 - 2f ( \alpha \eta - \alpha' \eta') ) \; \Big] \eeq
Now, the above equation still has the word "want" in it, and for a good reason. The change that we have "asked" for in the above equation is discrete; but we would like to accomplish it by continuous means. This, of course, should be possible since the above change is to take place during the "short" time interval; the latter has finite duration. Now, suppose we have the "desired" equation
\beq F_1 \rightarrow^{\rm want} F_2 \eeq
where $F_1$ and $F_2$ are some \emph{scalar} fields. In this case, we would like $F_1$ to increase if it is smaller than $F_2$, and we would like it to decrease if it is larger than $F_2$. It is easy to see that this can be accomplished by postulating an equation
\beq \frac{\partial F_1}{\partial t} = \mu (F_2 - F_1) \eeq
where the constant $\mu$ has to be very large, so that $F_1$ "catches up" with $F_2$ \emph{before} $F_2$ "has time" to change. Now, we can get a little bit closer to the situation of our interest, and replace $F_1$ and $F_2$ with $B_{\mu \nu}$ and $C_{\mu \nu}$, respectively. Thus, if we have a "desired" equation
\beq B_{\mu \nu} \rightarrow^{\rm want} C_{\mu \nu} \eeq
it can be accomplished by postulating dynamics
\beq \frac{\partial B_{\mu \nu}}{\partial t} = \mu (C_{\mu \nu} - B_{\mu \nu}) \eeq
Now, to get even closer to the case of our interest, suppose our "desired" dynamics is 
\beq f (\chi - \chi_0) \approx 1 \Rightarrow^{\rm want} [B_{\mu \nu} \rightarrow C_{\mu \nu}] \eeq
In this case, we would like to have the dynamics that we have just proposed for the case $f (\chi - \chi_0) \approx 1$, and, at the same time, we would like to "freeze" the dynamics of $B_{\mu \nu}$ for the case $f (\chi - \chi_0) \approx 0$. In light of the fact that $f$ is an approximation to step function, \emph{any} equation would be appropriate, as long as it does what we want for the special cases of $f$ being close to either $0$ or $1$. Thus, in particular, our goal is accomplished by the dynamics 
\beq \frac{\partial B_{\mu \nu}}{\partial t} = \mu f(\chi - \chi_0) (C_{\mu \nu} - B_{\mu \nu}) \eeq
Finally, we recall that in our case, the desired equation has been 
\beq  f(\chi - \chi_0) \approx 1 \Rightarrow^{\rm want} [ B_{\mu \nu} \rightarrow  \lambda A_{\mu \nu} (1 - 2f ( \alpha \eta - \alpha' \eta') ) ] \eeq
This means that we can use a substitution
\beq C_{\mu \nu} =  \lambda A_{\mu \nu} (1 - 2f ( \alpha \eta - \alpha' \eta') ) \eeq
which will lead us to our final, deterministic, dynamical equation
\beq \frac{\partial B_{\mu \nu}}{\partial t} = \mu f(\chi - \chi_0) (\lambda A_{\mu \nu} (1 - 2f ( \alpha \eta - \alpha' \eta') )- B_{\mu \nu} ) \eeq

\subsection*{5. Discussion and outlook}

In this paper we have argued that it is possible to include gravity into Bohm's Pilot Wave model, based on the following scheme:

a) Propose flat space version of Pilot Wave model that we agree upon

b) Modify the above version for the situation of curved spacetime, with known curvature

c) Come up with a modification of \emph{classical} gravity, so that the \emph{classical} dynamics has exact solutions for non-conserved energy sources

d) Use the behavior of beables that we obtained in part b as a "non-conserved" source of subsequent evolution of metric

In light of the fact that quantum field theory in curved space time is a lot less controversial than actual quantum theory of gravity, we have assumed that we agree on the way to carry out steps "a" and "b". We claimed that the biggest difficulty sits within the step "c" and, once that "classical" problem is solved, the step "d" (which is the ultimate goal of the theory) goes through automatically. 

In the last section we have proposed one specific example of a way to carry out the step "c". Unfortunately, however, it is easy to see that the specific model that we have proposed is extremely unnatural. Basically, we have simply came up with some "mechanism" of randomly varying metric and repeatedly "checking out" during what we call "short intervals" as to whether or not we are doing it "right", and then "correcting ourselves" if we do not. From strictly mathematical point of view this is still a well defined, deterministic, dynamics. But, at the same time, the complexity of the dynamics proposed might not pass Ocam's razor. 

Apart from this, the theory of gravity proposed in this paper does not respect general relativistic covariance: in particular, we had to often use \emph{ordinary} derivatives in stead of covariant ones, and we had to also use absolute time. Our justification had been that we have to make piece with these notions in order to be able to talk about Pilot Wave model, to begin with. So since we have already violated covariance, we might as well do it some more. I believe, however, that the theory would look much better if we were to violate the covariance \emph{only} when we have to. In other words, it would be far more desirable to openly couple non-covariant quantum mechanics with completely covariant gravity. Unfortunately, at this point we do not know how to do that. 

The main reason why we haven't proposed more natural dynamics is simply that, at this point in time, we do not know how. However, for the future work it is crucial to try to find such dynamics. The dynamics that is being proposed in this paper should not be a candidate for a final answer. It is simply an example only meant to convince the reader that "classical" version of gravity advocated in this paper is "possible". Now, comming up with "better" dynamics has to, inevitably, involve the answers to the following questions:

A. Even if the new dynamics allows for the violation of energy momentum conservation, one has to make sure that we do not have any other, less obvious, "conservation law". Of couse, by Noether's Theorem, we would have some \emph{physical} conservation law. But we should make sure that the latter is not purely \emph{mathematical}. That is, we should not be able to derive any of the "conservation laws" based on Bianchi identities or any other purely mathematical theorems. 

B. We have to prove that, in fact, the dynamics for gravity has a solution regardless of the behavior of energy momentum tensor. The answer to this question is \emph{not} the same as the answer to the previous one. For example, if we go back to the "standart" situation where everything is conserved, it is not obvious that Einstein's equation does, in fact, have solutions. Of course, there are theorems that state that it does. But someone, somewhere, had to prove these theorems, first. Therefore, similar work has to be re-done for non-conserved case. 

Another issue that comes to mind is that there is, already, a movement working on modified theories of gravity. The goal of that movement, of course, is very different: they are trying to come up with cosmological models that do not have dark matter. In light of this, it might be intersting to consider "joining efforts" with cosmologists and try to find a modified theory of gravity that serves two purposes at the same time. At the moment, I don't have enough knowledge to say whether or not any of the existing modified theories of gravity allow for energy momentum non-conservation. But it might be interesting for the future research to see iif that is the case and, if not, to find out whether there are any other models that can be proposed that would achieve this purpose.

\end{document}